\documentclass[aps,prl,twocolumn,superscriptaddress,showpacs]{revtex4-1}

\usepackage{graphicx}

\begin{document}


\title{Universality of dispersive spin-resonance mode in superconducting BaFe$_2$As$_2$}


\author{C. H. Lee}
\affiliation{National Institute of Advanced Industrial Science and Technology (AIST), Tsukuba, Ibaraki 305-8568, Japan}
\affiliation{Transformative Research-Project on Iron Pnictides (TRIP), JST, Chiyoda, Tokyo 102-0075, Japan}

\author{P. Steffens}
\affiliation{Institut Laue Langevin, 6 Rue Jules Horowitz BP 156, F-38042 Grenoble CEDEX 9, France}

\author{N. Qureshi}
\affiliation{$I\hspace{-.1em}I$. Physikalisches Institut, Universit\"at zu K\"oln, Z\"ulpicher Str. 77, D-50937 K\"oln, Germany}

\author{M. Nakajima}
\author{K. Kihou}
\author{A. Iyo}
\author{H. Eisaki}
\affiliation{National Institute of Advanced Industrial Science and Technology (AIST), Tsukuba, Ibaraki 305-8568, Japan}
\affiliation{Transformative Research-Project on Iron Pnictides (TRIP), JST, Chiyoda, Tokyo 102-0075, Japan}

\author{M. Braden}
\affiliation{$I\hspace{-.1em}I$. Physikalisches Institut, Universit\"at zu K\"oln, Z\"ulpicher Str. 77, D-50937 K\"oln, Germany}




\begin{abstract}
Spin fluctuations in superconducting BaFe$_2$(As$_{1-x}$P$_x$)$_2$
(x=0.34, $T_c$ = 29.5 K) are studied using inelastic neutron
scattering. Well-defined commensurate magnetic signals are
observed at ($\pi$,0), which is consistent with the nesting vector
of the Fermi surface. Antiferromagnetic (AFM) spin fluctuations in the
normal state exhibit a three-dimensional character 
reminiscent of the AFM order in nondoped
BaFe$_2$As$_2$. A clear spin gap is observed in the
superconducting phase forming a peak whose energy is 
significantly dispersed along the c-axis.  
The bandwidth of dispersion becomes larger with approaching the AFM ordered phase 
universally in all superconducting BaFe$_2$As$_2$, 
indicating that the dispersive feature is attributed to three-dimensional AFM correlations.  
The results suggest a strong relationship between the magnetism and superconductivity.  
\end{abstract}

\pacs{74.70.Xa, 75.40.Gb,78.70.Nx}


\maketitle

Magnetism is considered to play a crucial role on the appearance
of superconductivity in iron-based superconductors
\cite{hirschfeld}. To verify this assumption, the relationship
between superconductivity and spin fluctuations has been studied
intensively. 
Carrier-doped superconducting $A$Fe$_2$As$_2$ ($A$ = Ba, Sr and Ca) 
is one of the systems which have been well studied using inelastic neutron scattering 
\cite{Christianson2008,Pratt2011,Park2010,Luo,CHLee2011,Chi2009,Li2009,Wang,Lipscombe,Liu,Christianson2009,Inosov,Lumsden2009,Zhang2011,Pratt2010,Zhang2013,Steffens} 
due to the availability of sizable single crystals.  
Spin fluctuations are observed around ($\pi$,0) [(0.5,0.5,L) in tetragonal notation] in the system, 
where their peak positions are determined by the
topology of the Fermi surface (FS) \cite{Park2010,Pratt2011,Luo,CHLee2011}. 
They exhibit a gap structure in the superconducting phase with
remarkable enhancement of the magnetic signal at a specific energy
\cite{Christianson2008,Christianson2009,Inosov,Lumsden2009,Zhang2011,Park2010,Chi2009,Li2009,Wang,Lipscombe,Liu,Pratt2010,Zhang2013,Steffens}.

The interpretation of the enhancement depends on the superconducting gap symmetry.  
For a s$_\pm$-wave gap, the enhancement is argued to be a spin resonance \cite{Mazin2008,Kuroki2008}.  
In contrast, for a s$_{++}$-wave gap, it is formed due to 
the absence of inelastic quasiparticle scattering just on the gap \cite{Onari2010,Onari2011}. 
Thus, it is essential to clarify the origin of the intensity enhancement to determine the superconducting gap symmetry.
In this paper, we refer to the intensity enhancement as resonance for convenience, 
although the interpretation of the enhancement is still controversial. 

The resonance characteristics of iron-based superconductors are complicated. 
In Co- and Ni-doped BaFe$_2$As$_2$, 
they depend on the $L$ value
\cite{Chi2009,Li2009,Wang,Park2010,Pratt2010} and exhibit a pronounced spin-space anisotropy \cite{Lipscombe,Zhang2013,Steffens}.  
The resonance energies are larger for even values of $L$ 
than for odd values.  
Polarized neutron scattering measurements in optimum Co-doped BaFe$_2$As$_2$ have
revealed an additional sharp peak at an energy of 1.8$k_BT_c$,
which is considerably below the resonance observed in unpolarized
neutron-scattering measurements \cite{Steffens}.  
Here, $k_{\rm B}$ denotes the Boltzmann constant.  
The origin of the anisotropies is still unexplained 
although clear understanding of the resonance is important 
for clarifying the mechanism of Cooper pair formation. 
Therefore, in the present paper, we try to reveal the origin in particular of the dependence on $L$.  

Two different interpretations have been proposed 
to explain the dispersive feature of the resonance.  
One has attributed it to the c-axis dependence of the superconducting gap 
under the assumption that the resonance energy is associated with the superconducting gap value \cite{Chi2009,Wang}.  
The other has argued that it arises from interlayer spin correlations 
based on the random-phase approximation (RPA) calculations 
where the resonance results from the formation of spin excitons \cite{Pratt2010,Eschrig}.  
Stronger spin correlation can cause lower resonance energies in this model.  

To solve the problem, it is efficient to compare the resonance in samples 
whose superconductivity is induced by different methods 
resulting in nonequivalent superconducting gap anisotropy.  
For this purpose, $A$Fe$_2$As$_2$ is a suitable system 
because superconductivity can be controlled in several ways, namely: 
electron \cite{Sefat} or hole doping \cite{Rotter}, external pressure \cite{Alireza},
and chemical pressure induced by P-doping at the As site
\cite{Jiang}.  
Although intensive studies on spin fluctuations have been carried out by inelastic neutron
scattering using single crystals of carrier-doped superconducting $A$Fe$_2$As$_2$, 
studies on samples whose superconductivity is induced
by other methods have been restricted to powder samples \cite{Ishikado}.
We therefore conducted inelastic neutron
scattering measurements on single crystals of superconducting BaFe$_2$(As,P)$_2$,
where the superconductivity occurs without carrier doping. We
found that the resonance energy depends strongly on $L$, indicating the
universality of its dispersive feature in the superconducting phase.

\begin{figure}
\includegraphics[width=\columnwidth]{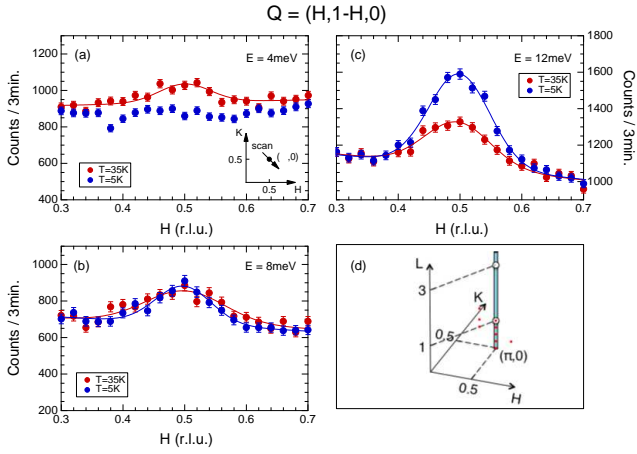}
\caption{\label{hk0} (a)-(c) Constant-energy scans of BaFe$_2$(As,P)$_2$ in the ($H$,$K$,0) zone above and below $T_c$ for (a) $E$ = 4 meV, (b) $E$ = 8 meV, and (c) $E$ = 12 meV.
The scan trajectory is indicated by the arrow in panel (a).  Solid lines indicate Gaussian fits.
(d) Schematic illustrations of magnetic peak positions forming a rod structure in BaFe$_2$(As,P)$_2$.
Open circles at (0.5,0.5,$L$) with odd $L$ describe the peak positions of AFM long-range order in nondoped BaFe$_2$As$_2$.  
Dots describe measured positions in constant-$\bf Q$ scans shown in Fig. \ref{intensity-kai}.}
\end{figure}

Single crystals of BaFe$_2$(As$_{1-x}$P$_x$)$_2$ were grown by the
self-flux method, which is described in detail elsewhere
\cite{Nakajima2012a}. Approximately 260 tabular-shaped single
crystals ($\sim$0.06 cm$^3$) taken from the same batch were
co-aligned on thin Al sample holders for inelastic neutron
scattering measurements. The total mosaic spreads of the
co-aligned samples had full widths at half maximum of $\sim$2.5$^\circ$ and $\sim$3.5$^\circ$ 
in the (H,H,L) and (H,K,0) scattering
planes used in our experiment, respectively. The $T_c$ of the
single crystals was determined to be 29.5 K from the temperature
dependence of the zero-field-cooled magnetization using a SQUID
magnetometer (Quantum Design MPMS) under a magnetic field of 10 Oe
parallel to the c-axis. The P content in the single crystals was
determined to be x = 0.34 based on the $T_c$ value and c-axis lattice
parameters obtained from X-ray diffraction \cite{Nakajima2012a}.

Inelastic neutron scattering measurements were conducted using the
triple-axis spectrometer IN8 at the Institute Laue-Langevin in
Grenoble, France. The final neutron energy was fixed at $E_{f}$ =
14.7 meV by using double-focusing pyrolytic graphite (PG) crystals
as a monochromator and analyzer. No collimator was used to
maximize the intensity. A PG filter was inserted to remove
higher-order neutrons. A He cryostat was used to cool the samples down
to 5 K.

\begin{figure}
\includegraphics[width=\columnwidth]{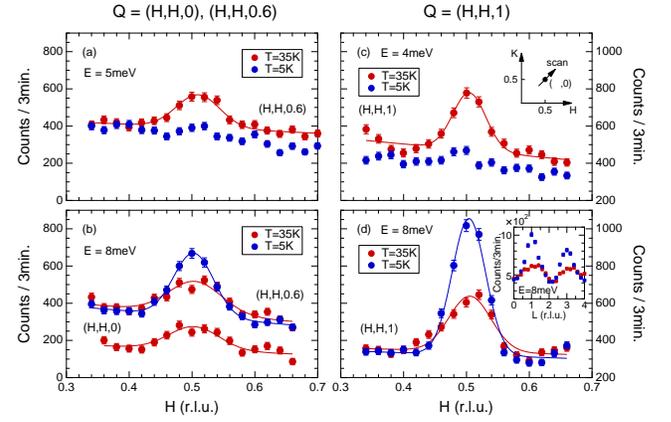}
\caption{\label{hhl} Constant-energy scans of BaFe$_2$(As,P)$_2$
in the ($H$,$H$,$L$) zone above and below $T_c$. (a,b) $\bf q$
scans at $L$ = 0 and 0.6 for (a) $E$ = 5 meV and (b) $E$ =
8 meV. The spectrum at $L$ = 0 is shifted down by 200 counts to
facilitate viewing. (c,d) $\bf q$ scans at $L$ = 1 for (c)
$E$ = 4 meV and (d) $E$ = 8 meV. The scan trajectory is indicated
by the arrow in panel (c).  Solid lines indicate Gaussian fits.
The inset in panel (d) shows the $L$ dependence of the magnetic peak
intensities above and below $T_c$ for $E$ = 8 meV.}
\end{figure}

\begin{figure}
\includegraphics[width=\columnwidth]{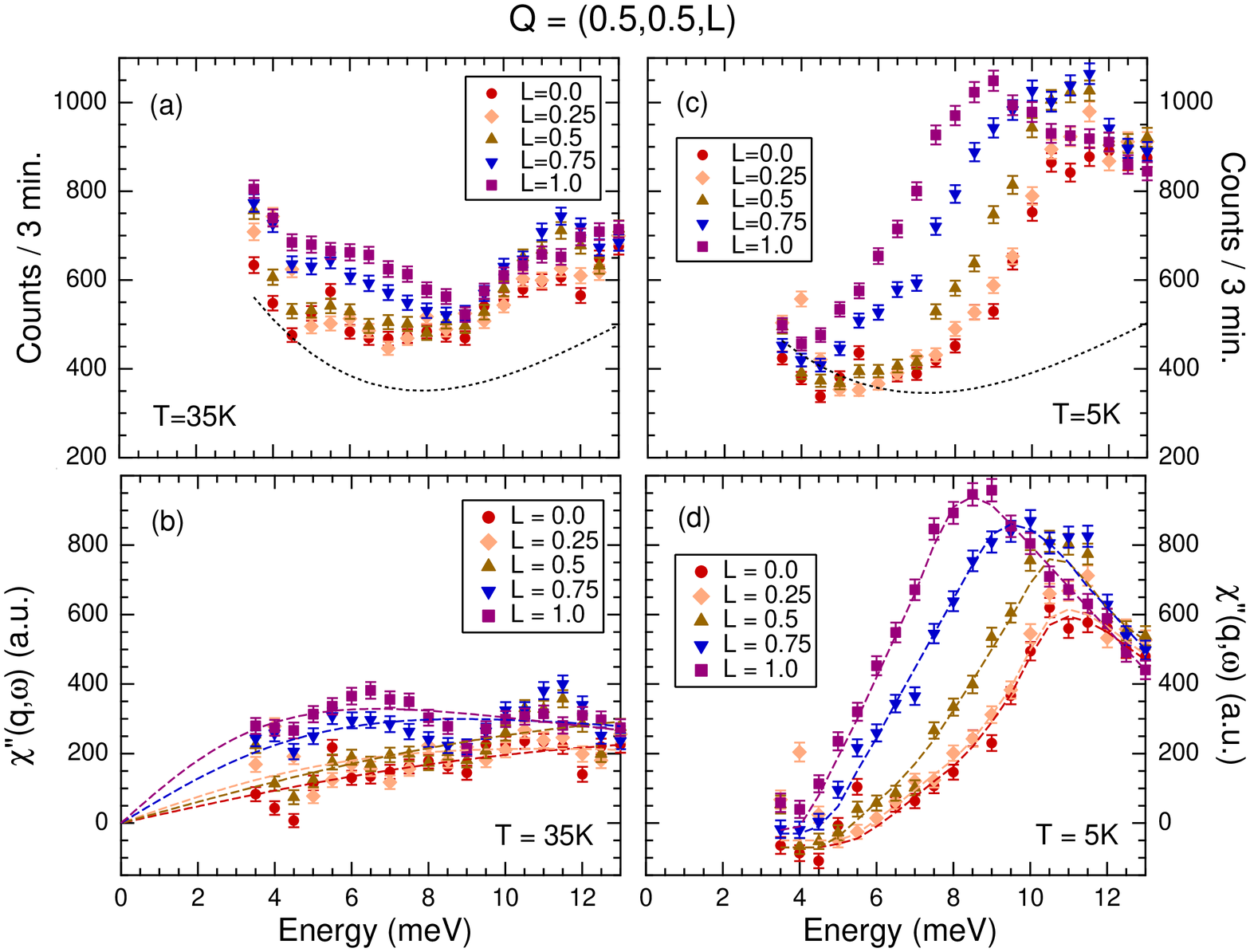}
\caption{\label{intensity-kai} (a,c) Constant-$\bf Q$ scans at
$\bf Q$ = (0.5,0.5,$L$) for several $L$ values (a) above and (c)
below $T_c$. Dotted lines describe the background determined by
averaging the background at $\bf Q$ = (0.38,0.38,1),
(0.38,0.38,1.25), (0.38,0.38,1.65), and (0.62,0.62,0).  Measured $\bf Q$ positions are described in Fig. \ref{hk0} (d).  (b,d)
Energy spectra of $\chi^{\prime \prime}({\bf q},\omega)$ (b) above
and (d) below $T_c$. These were obtained by subtracting the
background and by correcting for the monitor higher-order contamination and for
the Bose factor. The dashed lines in panel (b) are fits obtained
using Eq. (1), and those in panel (d) are guides to the eye.}
\end{figure}

Figures \ref{hk0}(a-c) show $\bf q$ scans measured in the [1,-1,0]
direction of the ($H$,$K$,0) zone at $T$ = 35 and 5 K.
Well-defined commensurate peaks are observed at $\bf Q$ =
(0.5,0.5,0) with a spin gap opening below $T_c$. At $E$ = 4 meV,
the magnetic intensity vanishes at $T$ = 5 K due to the spin gap.
At $E$ = 8 meV, the peak intensity remains constant upon cooling,
whereas it increases remarkably at $E$ = 12 meV.

Commensurate peaks are also observed in the $\bf q$-spectra measured in
the [1,1,0] direction of the ($H$,$H$,$L$) zone for integer and non-integer $L$
values (Fig. \ref{hhl}). These commensurate peaks indicate that
magnetic peaks form a rod structure along the c-axis at $\bf Q$ =
(0.5,0.5,$L$) [Fig. \ref{hk0}(d)]. Spin gaps are also
observed at $L$ = 0.6 and 1 below $T_c$ [Figs. \ref{hhl}(a,c)].
An enhancement of the peak intensity below $T_c$ is observed at $E$ = 8
meV for $L$ = 0.6 and 1 [Fig. \ref{hhl}(b,d)]. The peak
intensities strongly depend on the $L$ value at $E$ = 8 meV for both
$T$ = 5 and 35 K [Figs. \ref{hhl}(b,d)].  
They exhibit maximum intensity 
at odd values of $L$ [inset of Fig. \ref{hhl}(d)], where the
long-range antiferromagnetic (AFM) order in nondoped
BaFe$_2$As$_2$ forms magnetic Bragg peaks in elastic scattering.  

The energy dependence of the magnetic scattering at $T$ = 35 and 5 K was
measured at several $L$ values with the background determined
at the sides of the magnetic rod [Figs. \ref{intensity-kai} (a,c)]. The
dynamical magnetic susceptibility $\chi^{\prime \prime}({\bf
q},\omega)$ was obtained by multiplying the net intensity by
$[1-\exp(-\hbar\omega/k_{\rm B}T)]$ after correcting for
higher-order components in the incident beam monitor \cite{Shirane-book} [Figs.
\ref{intensity-kai} (b,d)].  $\chi^{\prime \prime}({\bf q},\omega)$ at $T$
= 35 K for $L$ = 0, 0.25 and 0.5 increases monotonically with
increasing energy and tends to saturate, as is
usually observed in correlated spin systems without magnetic
long-range ordering. Additional signals were observed below $E$ =
9 meV at $L$ = 0.75 and 1, 
in agreement with the $L$ dependence of the intensity
at E = 8 meV shown in the inset of Fig. 2(d).
Magnetic correlations in BaFe$_2$(As$_{0.66}$P$_{0.34}$)$_2$
possess hence a three-dimensional character reflecting
the fully  three-dimensional AFM order observed in nondoped
BaFe$_2$As$_2$. We verified that there is no detectable elastic magnetic
signal appearing at $\bf Q$ = (0.5,0.5,1)
above $T$ = 5 K, confirming that there is no sizeable static AFM
order in BaFe$_2$(As$_{0.66}$P$_{0.34}$)$_2$.

The energy dependence of $\chi^{\prime \prime}({\bf q},\omega)$ in the normal state was fitted
using a phenomenological function applicable to correlated spin systems in Fermi liquids without magnetic long-range ordering,

\begin{equation}
\chi^{\prime \prime}({\bf q},\omega) = \chi_{0}\frac{\Gamma\hbar\omega}{\Gamma^2+(\hbar\omega)^2} \ \ ,
\end{equation}

\noindent where $\chi_0$ represents the strength of the AFM
correlation, and $\Gamma$ is a damping constant. Fitting results are
shown by the solid lines in Fig. \ref{intensity-kai} (b).  
The evaluated $\Gamma$ decreases
from 20.5$\pm$7.5 meV to 6.5$\pm$1 meV when $L$ varies from 0 to 1. 
The fitted lines at $L$ = 0, 0.25 and 0.5 reproduce the data reasonably well, 
whereas those at $L$ = 0.75 and 1 reproduce the data only moderately well 
because the additional feature below $E$ = 9 meV cannot be described appropriately.  
Thus, the data at $L$ = 0.75 and 1 seem to be beyond the framework of equation (1).  
This could be because the sample is close to AFM long-range ordering.

\begin{figure}
\includegraphics[width=\columnwidth]{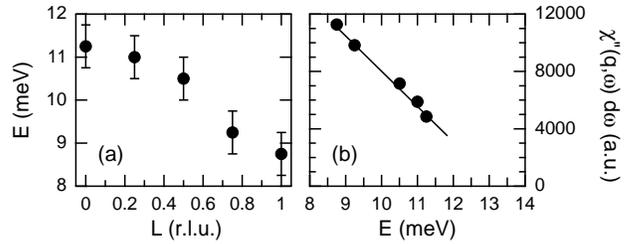}
\caption{\label{dispersion} (a) $L$ dispersion of resonance energy
defined as the intensity maximum in the energy spectra shown in
Fig. 3(d).  (b) Resonance energy dependences of energy-integrated $\chi^{\prime \prime}({\bf q},\omega)$ 
over $E$ = 3.5 to 13 meV at T = 5 K.}
\end{figure}

$\chi^{\prime \prime}({\bf q},\omega)$ in the superconducting 
state at T = 5 K exhibits apparent spin gap structures at all values of $L$ 
with a clear peak referred to as resonance [Fig. \ref{intensity-kai} (d)].  
The resonance energy and intensity strongly depend on $L$. 
The additional low-energy peak observed by polarized inelastic neutron scattering 
in Co-doped BaFe$_2$As$_2$ below $T_c$ \cite{Steffens} 
can also be included in the present $\chi^{\prime \prime}({\bf q},\omega)$.  
By analogy, the expected peak energy would be $E$ = 4.6 meV.  
In fact, subtle humps might be seen in the low energy region, 
although the energy is slightly higher than the expected one.  
Our experiment is, however, unable to fully separate the humps from
the main peak without using polarized neutrons.  
In the following, we will focus on
the main response whose intensity maxima can be unambiguously
determined due to its sufficiently high intensity, 
despite the overlap with the low-energy shoulders.  

Figure. \ref{dispersion} (a) shows the dispersion of the resonance.  
The observed resonance energies correspond to $2\Delta/k_BT_c$ = 3.4 - 4.4.  
The energy is lower for odd values of $L$, where magnetic signals of 
three-dimensional character are found in the normal state.  
The energy-integrated $\chi^{\prime \prime}({\bf q},\omega)$ up to $E$ = 13 meV at T = 5 K 
decreases linearly with increasing resonance energy [Fig. \ref{dispersion} (b)].  
The linear relationship is qualitatively consistent with the RPA calculations 
where the resonance is interpreted as the formation of spin excitons \cite{Park2010,Eschrig,Millis}.  
It is predicted that the energy of resonance decreases and intensity increases with increasing an exchange constant.  
Because the calculations were applied to d-wave superconductors, 
they should be extended to deal with the s-wave gap to explain the present results.  
The required assumption is that the onset of the particle-hole continuum should be independent of $L$.  
According to angle-resolved photoemission spectroscopy, 
superconducting gaps of BaFe$_2$(As$_{1-x}$P$_x$)$_2$ depend on the c-axis wave vector 
and exhibit a nodal gap structure at the Z-point on the outmost hole FS \cite{Zhang-ARPES}.  
However, the nodal gap cannot directly be sensed in our experiment.  
First the size and the orbital character do not match with the counterpart electron FS.  
In addition, the inelastic neutron scattering experiment integrates all processes 
between the nested FS'es which are gapped in the superconducting state.  
In particular, it averages the distribution of gap values along the c-axis.  
Isotropic superconducting gaps along the c-axis are, thus, not required. 

\begin{figure}
\includegraphics[width=\columnwidth]{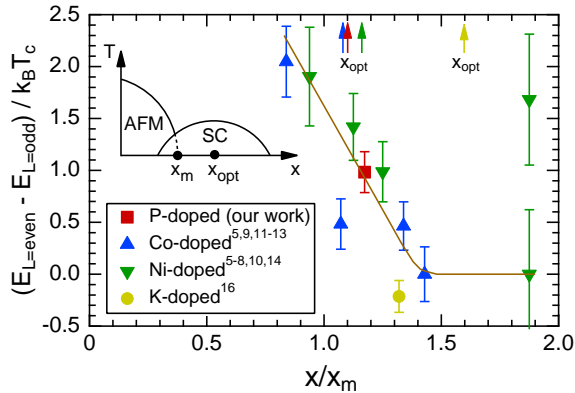}
\caption{\label{resonance} Doping dependence of the bandwidth ($E_{\rm L=even} - E_{\rm L=odd}$) of the resonance dispersion along $L$ 
for various iron-based superconductors \cite{Pratt2010,Steffens,Inosov,Park2010,Lumsden2009,Chi2009,Li2009,Wang,Lipscombe,Liu,Zhang2011}.
Energies are normalized by $k_{\rm B}T_c$ and the doping levels are normalized by the levels where AFM disappears ($x_{\rm m}$).
The $x_{\rm m}$ values are assumed to be 29 \%, 5.6 \%, 4 \%, and 25 \% for P-, Co-, Ni-, and K-doped samples, respectively.
The optimum doping level (x$_{\rm opt}$) is depicted by arrows.
The solid line is a guide to the eye.}
\end{figure}

The doping dependence of the bandwidth ($E_{\rm L=even} - E_{\rm L=odd}$) of the resonance dispersion along $L$ is
analyzed for various iron-based superconductors in Fig. \ref{resonance}.  
For comparison, energies are normalized by
$k_{\rm B}T_c$, and the doping levels are normalized by
compositions where AFM ordering disappears ($x_{\rm m}$).  
Values of $x_{\rm m}$ were determined from each phase diagram by
extrapolating their N$\rm \acute{e}$el temperatures \cite{Nakajima2012a,Kasahara,Chu,Canfield,Avci,Rotter-BaK-2008b}. 
In fact, the two samples which have values of $x/x_{\rm m}$ less than 1
exhibit AFM long-range ordering \cite{Wang,Pratt2010}, whereas the others do not, which
justifies the present estimation of $x_{\rm m}$. 
Although the estimation is rough, 
it is sufficient to figure out the overall picture.  
As shown in Fig. \ref{resonance}, the normalized bandwidth 
increases with decreasing doping level following a universal
curve. 
This suggests that the energy dispersion has a common 
origin and that it is related to the AFM state. The dispersion 
has not been found in K-doped samples yet, which is likely due to
too high doping of the samples used in inelastic neutron
scattering experiments.  

Based on the results, the carrier concentration cannot be an essential factor for the
$L$ dispersion of the resonance as there is no electron
doping in P-doped samples. Also details of the FS nesting can be discarded,
because the topology of FS should be different between electron
and P-doped samples. Either the nodal or anisotropic structure of the superconducting
gap should also be irrelevant because 
samples with optimum electron doping exhibit a full gap \cite{Reid}, whereas optimum
P-doped samples exhibit a nodal gap \cite{Hashimoto-P}, although they show similar 
$L$ dispersion. 
The dispersion can rather be 
associated with three-dimensional AFM correlations, as it clearly
depends on the doping levels normalized by $x_{\rm m}$. A larger dispersion 
can be caused by a larger three-dimensional AFM correlation. In
fact, the present P-doped sample shows short-range
three-dimensional AFM correlation in the normal state and the resonance 
energy is low at a q-position where the AFM correlation is strong.
Electron-underdoped superconducting samples which show the largest
dispersion even exhibit three-dimensional AFM ordering
\cite{Wang,Pratt2010}. These results provide evidence that the $L$
dispersion is related to the three-dimensional character of AFM
correlations consistent with the RPA calculations, 
suggesting a strong relationship between magnetism and superconductivity.  
The resonance can be explained by the spin exciton model, 
which supports the s$_\pm$-wave gap.  

In summary, well-defined commensurate spin fluctuations have been
observed at ($\pi$,0) in optimum doped
BaFe$_2$(As$_{1-x}$P$_x$)$_2$.
A clear spin gap forms in the superconducting phase with resonance energies 
dispersing along the $c$ direction.  A lower resonance energy
is observed at (0.5,0.5,$L$ = odd). The dispersive feature of the
resonance can be attributed to the three-dimensional character
of AFM correlations and thus to the neighborhood of the static AFM
ordered phase, suggesting a strong relationship between magnetism and superconductivity. 

\begin{acknowledgments}
We wish to thank T. Tohyama and T. Yoshida for valuable discussions.
This study was supported by a Grant-in-Aid for Scientific Research
B (No. 24340090) from the Japan Society for the Promotion of
Science and by the Deutsche Forschungsgemeinschaft through SFB608.
\end{acknowledgments}

\end{document}